# Deep Recurrent Architectures for Seismic Tomography

Amir Adler[1], Mauricio Araya-Polo[2], Tomaso Poggio[1]

[1]Center for Brains, Minds and Machines (CBMM), MIT, Cambridge, MA, USA
[2]Shell International Exploration and Production Inc., Houston, TX, USA
Corresponding author: adleram@mit.edu

**Introduction**

In hydrocarbon exploration velocity model building (VMB) is a key step, where tomography and full waveform inversion (FWI) are the main tools. VMB main product is an initial model of the subsurface that is subsequently used in seismic imaging and interpretation workflows. This paper introduces novel deep recurrent neural network architectures for VMB. Existing solutions rely on numerical solutions of the wave equation, which is highly demanding on computing and domain experts' time. Recently, a non-recurrent convolutional neural network (CNN)-based architecture was proposed (Araya-Polo et al, 2018), which enables significant reduction in computation time, as compared for example to FWI, directly producing a velocity model from raw seismic data.

Recurrent neural networks were designed for processing sequential data (Greff et al, 2017), which is an inherent characteristic of seismic traces, that are fundamentally time series of spatially sampled acoustic waves recordings. Our investigation includes the utilization of basic recurrent neural network (RNN) cells, as well as Long Short Term Memory (LSTM) and Gated Recurrent Unit (GRU) cells. Performance evaluation using a dataset of 12,000 velocity models and associated synthetic gathers reveals that salt bodies are consistently predicted more accurately by GRU and LSTM-based architectures, as compared to non-recurrent architectures. In addition, the RNN-based architecture enables 78% reduction in the total number of network coefficients, with the consequential memory savings, all this while preserving reconstructed model quality.

**Deep Recurrent Neural Network Architectures for Seismic Tomography**

Deep learning has recently emerged as a promising approach for seismic related tasks, for example by (Zhang et al, 2014), (Wang et al, 2018) and (Das et al, 2018). A non-recurrent convolutional neural network (CNN)-based architecture was proposed by (Araya-Polo et al, 2018), which processes raw seismic gathers (or their features) and outputs a prediction of the velocity model, as depicted in Figure 1. The network is trained by a data-driven approach (essentially solving a seismic inverse problem), in which a collection of training examples is generated, where each example includes a pair of a velocity model and associated seismic gathers. Gradient-based algorithms drive the network parameters learning, minimizing the predicted models and the ground truth misfit.

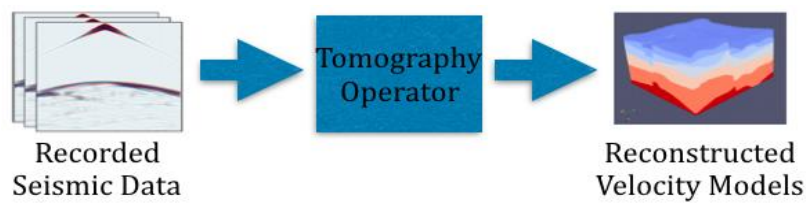

*Figure 1* Seismic Tomography workflow by a deep neural network.

Recurrent neural networks enable to process sequential data, by incorporating memory of previous hidden states. The simplest recurrent cell is the RNN, characterized by the following equation:

$$(1) \quad h_t = f(W_{hh}h_{t-1} + W_{hx}x_t),$$

Where $x_t$ the input at time $t$, $h_t$ is the hidden state at time $t$, $W_{hh}$ and $W_{hx}$ are the matrices of learnable weights, and $f(\bullet)$ is an element-wise activation function. The operation of the RNN cell along the time domain can be illustrated by unfolding the time axis as depicted in Figure 2. RNN cells were also proved to be equivalent to non-recurrent deep Residual Networks (Liao and Poggio, 2016), which have been demonstrated to provide excellent performance in computer vision tasks. Due to the

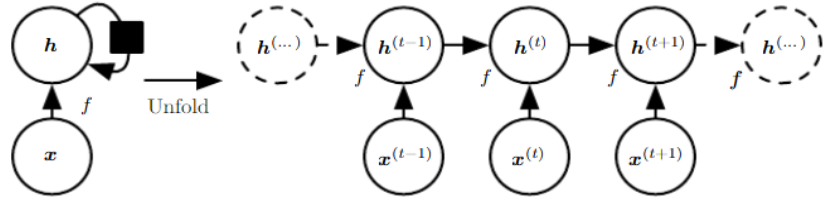

*Figure 2 Time unfolding of a recurrent neural cell (Goodfellow et al, 2016).*

vanishing and exploding gradients problems (Goodfellow et al, 2016) of RNNs, the more complex LSTM cell was proposed, which includes in addition to the hidden state, also a cell state with memory and three gates: forget, input, and output that control the flow of information within the cell depicted in Figure (3.a). The LSTM cell is characterized by equations (2.a)-(2.f):

(2.a) $f_t = sigmoid(W_f[h_{t-1}, x_t])$

(2.b) $i_t = sigmoid(W_i[h_{t-1}, x_t])$

(2.c) $o_t = sigmoid(W_o[h_{t-1}, x_t])$

(2.d) $g_t = \tanh(W_g[h_{t-1}, x_t])$

(2.e) $c_t = f_t \otimes c_{t-1} + i_t \otimes g_t$

(2.f) $h_t = o_t \otimes \tanh(c_t)$

(3.a) $z_t = sigmoid(W_z[h_{t-1}, x_t])$

(3.b) $r_t = sigmoid(W_r[h_{t-1}, x_t])$

(3.c) $c_t = \tanh(W_c[h_{t-1} \otimes r_t, x_t])$

(3.d) $h_t = (1 - z_t) \otimes c_{t-1} + z_t \otimes c_t$

The memory requirements of LSTM cells are four times larger than RNN, and it has been demonstrated to capture long term dependencies in the data (Greff et al, 2017). GRU is a reduced complexity version of LSTM with only two gates (reset and update), and it was demonstrated (Chung et al, 2014) to provide comparable sequence modelling performance as LSTM. The GRU cell is characterized by equations (3.a)-(3.d), and depicted in Figure 3(b).

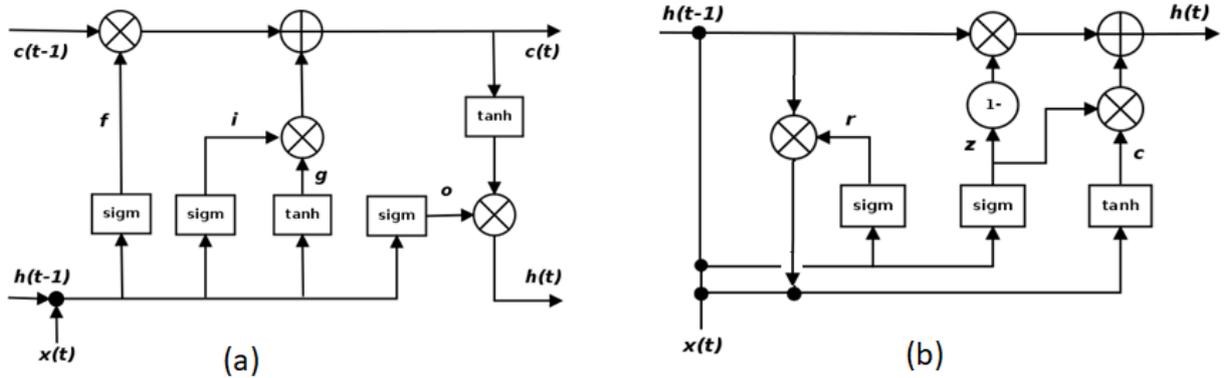

*Figure 3 LSTM (a) and GRU (b) cells. Element-wise multiplications are denoted by $\otimes$.*

The baseline Tomography non-recurrent CNN network (Araya-Polo et al, 2018) is composed of convolutional and fully-connected (FC) layers, as detailed in Table 1. Layers 1 and 2 serve as feature extractors, layers 3 and 4 reconstruct a 35 x 45 pixels image. Layers 6 - 9 serve as a super-resolution stage, which produces the final 70 x 90 pixels velocity models. The three recurrent networks architectures were constructed by removing layers 3 and 4 from the CNN network, and adding each of the corresponding recurrent cells RNN, LSTM or GRU, each configured with 512 hidden recurrent states. Note that the Maxpool layer is not required for the recurrent networks, since the data is entering sequentially into the recurrent cell.

| Network | Layer 1 | Layer 2 | Layer 3 | Layer 4 | Layer 5 | Layer 6 | Layer 7 | Layer 8 | Layer 9 |
|---|---|---|---|---|---|---|---|---|---|
| **CNN** | Conv | Maxpool | FC | FC | FC | Upscale | Conv | Conv | Conv |
| **RNN** | Conv | RNN | FC | Upscale | Conv | Conv | Conv | - | - |
| **LSTM** | Conv | LSTM | FC | Upscale | Conv | Conv | Conv | - | - |
| **GRU** | Conv | GRU | FC | Upscale | Conv | Conv | Conv | - | - |

*Table 1 Architectures of the evaluated deep networks.*

**Performance Evaluation**

We have generated a dataset of 12,000 velocity models with 4-8 layers, out of which 9,000 contain salt bodies. The velocity of the models ranges from 2000 to 4000[m/s], and the salt bodies from 4450 to 4550[m/s]. From each velocity model, multiple seismic gathers were generated by wave-based forward modelling. The dataset split is: 9,600 example pairs for training and a test-set of 2,400 example pairs, each pair consisting of a velocity model (the label) and seismic gathers (the data). Each network was implemented in TensorFlow, trained using the train-set over 50 epochs, mini-batch size of 100, and using the RMSprop optimizer with mean squared error (MSE) a loss function. Table 2 compares the various network architectures in terms of the number of coefficients and image reconstruction quality, over 2,400 test models, as measured using the structural similarity SSIM (Wang et al, 2004) and MSE. The results obtained by GRU have the lowest MSE and highest visual quality, as also illustrated in Figure 4. In addition, the RNN-based network achieves better structural similarity results than the CNN network, while saving over 78% of the coefficients.

| Network Type | CNN | RNN | LSTM | GRU |
|---|---|---|---|---|
| Number of coefficients in recurrent layer | 0 | **655,872** | 2,623,488 | 1,967,616 |
| Total number of coefficients | 7,182,728 | **1,557,896** | 3,525,512 | 2,869,640 |
| Percentage of coefficients vs. CNN network | 100% | **21.68%** | 49.08% | 39.95% |
| SSIM (averaged on 2,400 velocity models) | 0.8199 | 0.8210 | 0.8378 | **0.8414** |
| MSE (averaged on 2,400 velocity models) | 0.0018 | 0.0019 | 0.0014 | **0.0013** |

*Table 2 VMB image quality (higher SSIM values are better) and number of coefficients comparison.*

**Conclusions**

RNN were tailored for processing sequential data, and therefore are mostly suitable for processing seismic traces and their features. This work presents three VMB network architectures, based on RNN, LSTM and GRU cells. Performance evaluation reveals that on average the visual quality of velocity models prediction with the recurrent architectures is higher than non-recurrent architectures. In addition, salt bodies prediction is more accurate where subtle details are better resolved by the recurrent architectures, in particular with the LSTM and GRU networks. In terms of the total number of network coefficients, a significant reduction of up to 78% is achievable by utilizing recurrent architectures, which further motivates their usage for VMB. Future research includes architectures with multiple recurrent layers, and analysis of long term dependencies in seismic data, in order to fully understand the high quality results of the LSTM and GRU networks.

**References**

M. Araya-Polo, J. Jennings, A. Adler and T. Dahlke [2018]. Deep Learning Tomography, *The Leading Edge*, pp. 58-66, 37(1).

K. Greff, R. K. Srivastava, J. Koutník, B. R. Steunebrink and J. Schmidhuber [2017]. LSTM: A Search Space Odyssey, in *IEEE Trans. on Neural Networks and Learning Systems*, vol. 28, no. 10.

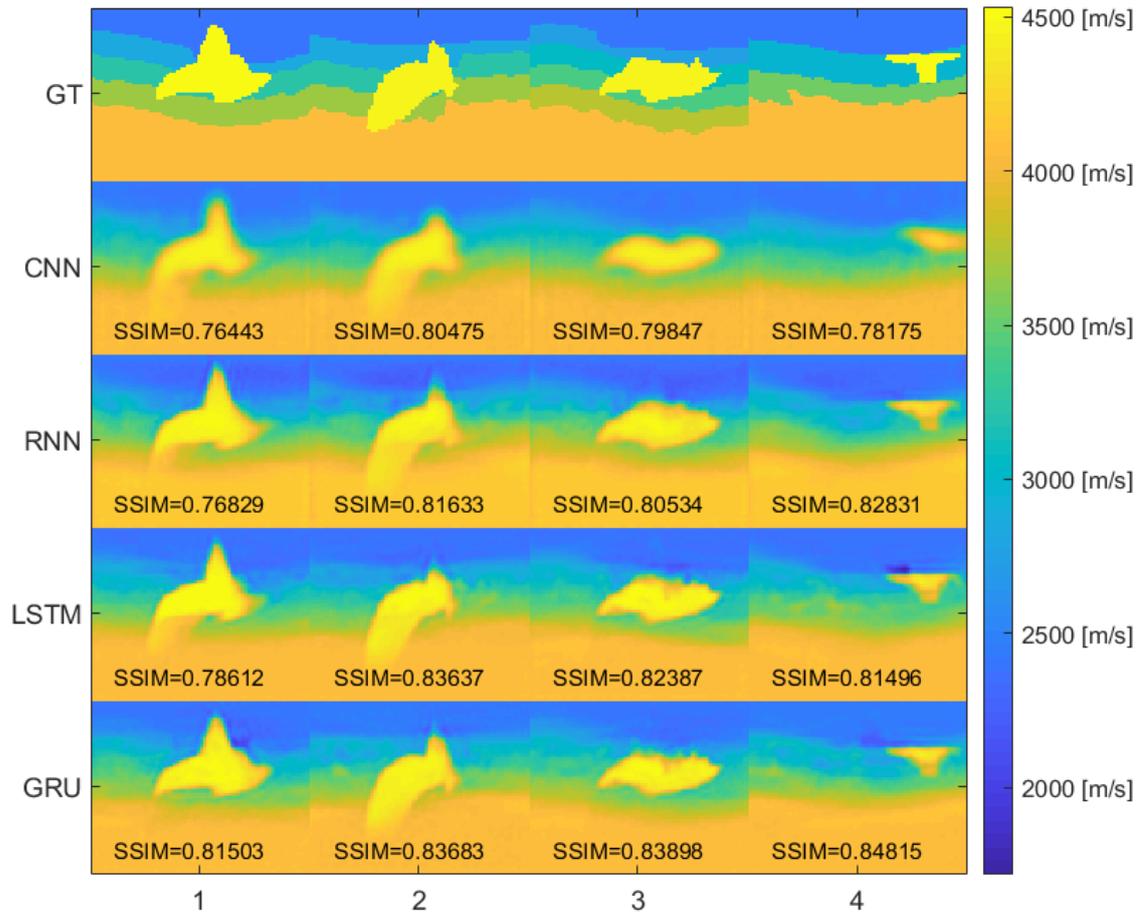

***Figure 4*** *VMB results of 4 models with salt bodies: 1st row ground truth (GT); 2nd row CNN (non-recurrent); 3rd row RNN; 4th row LSTM; and 5th row GRU results. The dimensions of each image are 70 x 90 pixels, representing depth (vertical axis) and lateral offset (horizontal axis).*

C. Zhang, C. Frogner, M. Araya-Polo and D. Hohl [2014]. Machine-learning Based Automated Fault Detection in Seismic Traces. 76th EAGE Conference and Exhibition.

B.Wang, N. Zhang and W. Lu. [2018]. Intelligent shot-gather reconstruction using residual learning networks. *SEG Technical Program Expanded Abstracts*, 2001-2005.

V. Das, A. Pollack, U. Wollner and T. Mukerji. [2018]. Convolutional neural network for seismic impedance inversion. *SEG Technical Program Expanded Abstracts*, 2071-2075.

I. Goodfellow, Y. Bengio and A. Courville [2016]. Deep Learning, MIT Press.

Q. Liao and T. Poggio [2016]. Bridging the Gaps between Residual Learning, Recurrent Neural Networks and Visual Cortex.

J. Chung and Ç. Glçehre and K. Cho and Y. Bengio [2014]. Empirical Evaluation of Gated Recurrent Neural Networks on Sequence Modelling, https://arxiv.org/abs/1412.3555

Z. Wang, A. C. Bovik, H. R. Sheikh and E. P. Simoncelli [2004]. Image quality assessment: from error visibility to structural similarity, in *IEEE Transactions on Image Processing*, vol. 13, no. 4, 2004.